\pdfoutput=1
\documentclass[journal=jctcce,manuscript=article]{achemso}

\usepackage{graphicx}
\usepackage{pdfpages}   % arXiv version: appends the Supporting Information PDF
\usepackage{amsmath,amssymb}
\usepackage{booktabs}
\usepackage{algorithm}
\usepackage{algpseudocode}
\usepackage{url}
\usepackage{xurl}   % r3c: allow line breaks inside long URLs
\SectionNumbersOn
\setkeys{acs}{articletitle=true}

\title{Foundation-Model Supervision of Distilled Potentials for Million-Atom Molecular Dynamics}

\author{Xiaoqing Liu}
\affiliation{Department of Mechanical Engineering, National University of Singapore, 117575, Singapore}
\author{Yangshuai Wang}
\email{yswang@nus.edu.sg}
\affiliation{Department of Mathematics, National University of Singapore, 10 Lower Kent Ridge Road, 119076, Singapore}
\author{Lei Shen}
\email{shenlei@nus.edu.sg}
\affiliation{Department of Mechanical Engineering, National University of Singapore, 117575, Singapore}

\abbreviations{MLIP, MD, DFT, MACE, NEP}

\begin{document}

\begin{abstract}
Machine-learned interatomic potentials distilled from foundation models enable fast large-scale molecular dynamics
(MD), but they can fail locally in chemical environments that their training data did not sample. System-averaged
error metrics and small random audits often miss such sparse or isolated failures. Here we use an exact equivariant
foundation model, MACE-MPA-0, to check every atom of million-atom production simulations. Reorganizing two-layer MACE
inference into a receiver-chunked three-pass schedule that keeps no per-edge intermediates resident reduces the
marginal memory per atom 340-fold, evaluates up to $2.86\times10^{7}$ atoms on a single H200 GPU, and supervises
$10^{6}$-atom trajectories in LAMMPS at 1.3\% of the run time. Distilled MACE students run four to nine times slower
than NEP89 with 0.95--1.28 times the teacher's force error against density functional theory (DFT). In molten
SiO$_2$, whole-system checks uncover a persistent tail of under-coordinated atoms hidden by the system-averaged error
and an isolated, unphysical O--O contact that a 1\% random audit misses 94\% of the time. On-the-fly retraining on
teacher-labelled 13-\AA{} clusters, with the force loss restricted to their inner 6~\AA, removes three quarters of the
failure tail and the kilobar-level pressure biases in SiO$_2$ and compressed water; the repair converges to the
teacher, not to DFT.
\end{abstract}

%% ====================================================================================================================
\section{Introduction}

Machine-learned interatomic potentials (MLIPs) bring \textit{ab initio} accuracy to molecular dynamics (MD) at a small
fraction of the cost of density functional theory (DFT).\cite{batzner2022nequip,musaelian2023allegro,batatia2022mace,zeng2023deepmd}
Two model families dominate large-scale simulations. Descriptor-based potentials such as NEP\cite{fan2021nep,fan2022gpumd}
and ACE\cite{drautz2019ace,lysogorskiy2021pace} evaluate a fixed set of per-atom many-body descriptors followed by a
small neural network or linear readout; their GPU implementations reach $\sim10^{7}$ atoms at $\sim10^{7}$
atom-steps\,s$^{-1}$ on one device, and the universal NEP89 model spans 89 elements.\cite{liang2026nep89} Equivariant
message-passing networks, and MACE\cite{batatia2022mace,batatia2025design} in particular, offer systematically higher
accuracy and transferability. Foundation models pretrained on MPtrj and Alexandria\cite{deng2023chgnet,schmidt2023alexandria,schmidt2024alexandria,barrosoluque2024omat24}
are increasingly used as drop-in replacements for DFT across the periodic table.\cite{batatia2025foundation} On a
single GPU, however, standard equivariant implementations run out of memory at $10^{4}$--$10^{5}$ atoms and are two
orders of magnitude slower per atom-step.

Knowledge distillation bridges this speed gap. Compact student potentials trained on foundation-model labels inherit
most of the teacher's accuracy at a fraction of its cost,\cite{gardner2024synthetic,gardner2025distillation} and
foundation models have recently been distilled into DeePMD, PiNet2 and NEP architectures, often combined with active
learning.\cite{lian2026distillation,sudhama2026pinnacle,zhang2026minimum} A distilled student is reliable only within
its training distribution, and its failures in production MD are typically local: a coordination defect, a stretched
bond or a compressed region where a handful of atoms experience badly wrong forces while the system-averaged error
remains small. Existing on-the-fly schemes guard against such failures with ensemble uncertainty estimates or reference
evaluations on small cells or sampled atoms.\cite{jinnouchi2019prb,jinnouchi2019prl,vandermause2020flare,zhang2020dpgen}
Whether checking every atom of a million-atom production simulation uncovers failures that sample-based audits miss,
and whether fine-tuning on carved local clusters repairs the running simulation, has not been tested in controlled
experiments.

Evaluating an equivariant foundation model on complete million-atom frames is the main obstacle. Most acceleration
efforts reduce constant prefactors or distribute memory over several GPUs: fused equivariant kernels
(cuEquivariance\cite{nvidia2024cuequivariance}), compiled inference,\cite{ansel2024pytorch2} the Kokkos-based ML-IAP
interface of LAMMPS,\cite{thompson2022lammps,smith2025mliapkokkos,johansson2025lammpskokkos} and spatial decomposition
across GPUs for Allegro and SevenNet.\cite{kozinsky2023scaling,park2024sevennet} Because these approaches retain
edge-level intermediates of all layers for reverse-mode force differentiation, the single-GPU memory per atom is
unchanged. Pruning message-passing layers and partitioning the graph fits a million atoms on one GPU but alters the
potential.\cite{kong2026pruning} Streaming tensor-product convolutions accumulate edge-local products directly into
receiver states and so avoid edge-sized intermediates within a single operator.\cite{chorosajev2026sobek} Concurrently
with this work, Symmetrix-XL\cite{fu2026symmetrixxl} combined streamed edges with tiled execution and
checkpoint-specific CUDA kernels for pretrained MACE models, reaching $1.12\times10^{7}$ atoms on an 80-GB A100 GPU in
LAMMPS with the same three sweeps as the schedule described here. Deployment-oriented equivariant architectures go
further: DPA4C keeps a single message-passing layer with tabulated distance-dependent edge functions and runs
$1.4\times10^{7}$ atoms on one H20 GPU faster than NEP89.\cite{li2026dpa4,li2026dpa4c} Like distilled students, such
fast universal models remain vulnerable where training data are sparse and benefit from exact whole-system checks.

\begin{figure*}[tbp]
\centering
\includegraphics[width=\linewidth]{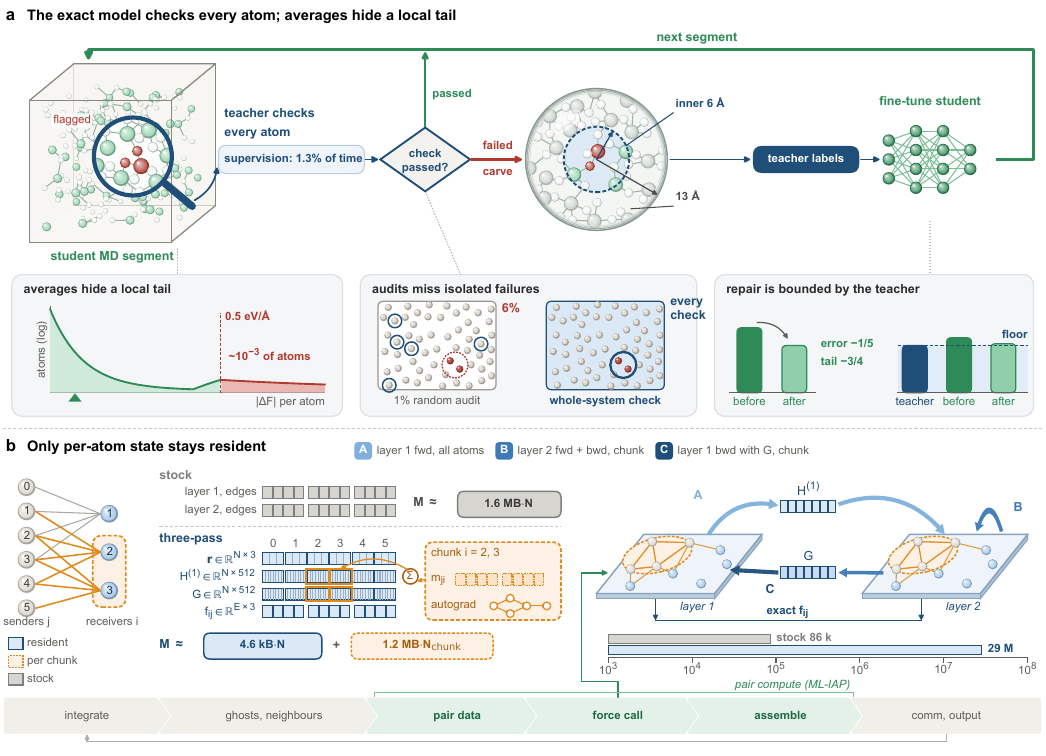}
\caption{Whole-system foundation-model supervision and memory-efficient three-pass evaluation. (a) Supervision
workflow, drawn for molten SiO$_2$. The distilled student drives MD segments while the exact MACE-MPA-0 teacher
recomputes every atom of dumped frames in LAMMPS (\texttt{rerun}); the supervision costs 1.3\% of the run time at one
check per 100~ps for $10^{6}$ atoms. A failed check carves 13-\AA{} clusters around the atoms with the largest errors,
labels them with the teacher in vacuum, and fine-tunes the student for 40--60 epochs with the force loss restricted to
the inner 6~\AA. Insets (schematic shapes, measured values): the averaged error hides a tail of $\sim10^{-3}$ of the
atoms; a 1\% random audit contains the isolated O--O contact with 6\% probability; retraining lowers the error towards
the teacher's own error against DFT. (b) Receiver-chunked three passes (A, layer-1 forward; B, layer-2 forward and
backward; C, layer-1 recomputation and backward) keep only per-atom arrays resident (4.57~kB per atom, against
1.55~MB for the stock evaluation), raising the capacity of one H200 from 86,016 to $2.86\times10^{7}$ atoms. Bottom
ribbon: one MD step in LAMMPS with the Kokkos ML-IAP interface; modified modules in green.}
\label{fig:overview}
\end{figure*}

Here we establish whole-system foundation-model supervision for large-scale MD and test in controlled experiments what
it detects and repairs (Figure~\ref{fig:overview}). Fast MACE students distilled from MACE-MPA-0 drive the trajectory,
while the exact teacher periodically checks complete frames of the running simulation, analogous to on-the-fly DFT
force fields in VASP\cite{jinnouchi2019prb,jinnouchi2019prl} and FLARE\cite{vandermause2020flare} but with an exact
foundation model in place of DFT. To make million-atom teacher evaluations feasible on one GPU, we reorganize two-layer
MACE inference into three receiver-chunked passes that keep no per-edge intermediates resident. The schedule is
algebraically exact, reduces the marginal memory per atom 340-fold, evaluates up to $2.86\times10^{7}$ atoms standalone
($8.96\times10^{6}$ inside LAMMPS) on one H200 GPU, and generalizes layer by layer to deeper NequIP-type networks
(SevenNet-0). Across molten SiO$_2$, liquid water and CuNi we show that whole-system checks capture both persistent
under-coordinated defect tails and isolated catastrophic contacts that averaged metrics and random audits miss, and
that fine-tuning on teacher-labelled 13-\AA{} clusters repairs these failures up to the accuracy of the teacher.

%% ====================================================================================================================
\section{Theory and Methods}

\subsection{Receiver-Chunked Three-Pass Evaluation of the Teacher}

\textbf{Origin of the memory bottleneck.} For $N$ atoms with on average $\bar n$ neighbors within the cutoff
$r_{\max}$, descriptor-based models (NEP) and equivariant message-passing models (MACE) both perform $O(N\bar n)$
operations, but their memory footprints differ. NEP reduces per-edge radial and angular descriptors immediately to
per-atom descriptors and recomputes edge derivatives during force assembly, so that its resident memory is $O(N)$ with
a small prefactor: 7.4~KiB per atom for NEP89 on an H200 GPU (measured). A MACE interaction layer instead forms
high-dimensional equivariant edge messages (radial-MLP weights $\times$ spherical harmonics $\times$ sender features)
and obtains forces by reverse-mode differentiation of the total energy. Every per-edge intermediate of both layers is
therefore kept in GPU memory until the backward pass: 1.55~MB per atom for MACE-MPA-0 on liquid water
($r_{\max}=6$~\AA, $\bar n=89.7$) with the stock LAMMPS ML-IAP path. Node feature width and arithmetic intensity are
properties of the model, whereas resident edge tensors and the full-system autograd graph are properties of the
evaluation schedule and can be removed (Figure~\ref{fig:overview}b).

\textbf{Three-pass formulation.} A two-layer MACE model writes the total energy as a sum over receivers $i$ with
neighbor sets $\mathcal N(i)$,
\begin{equation}
\begin{aligned}
E&=\sum_i \Big[E^{(0)}_{Z_i} + \varepsilon^{(1)}\big(h^{(1)}_i\big) + \varepsilon^{(2)}\big(h^{(2)}_i\big)\Big],\\
h^{(1)}_i &= \Phi_1\big(\{\mathbf r_{ij}\}_{j\in\mathcal N(i)}\big),\quad
h^{(2)}_i = \Phi_2\big(h^{(1)}_i,\{\mathbf r_{ij},h^{(1)}_j\}_{j\in\mathcal N(i)}\big).
\end{aligned}
\label{eq:energy}
\end{equation}
The layer-1 features $h^{(1)}_i$ (512 numbers per atom for MACE-MPA-0) are the only quantity that couples different
receivers across the two layers. With the accumulated upstream gradient
$G_j\equiv\sum_i\partial\varepsilon^{(2)}(h^{(2)}_i)/\partial h^{(1)}_j$, the chain rule splits the derivative of the
energy with respect to an edge vector $\mathbf r_{ij}$ into two terms,
\begin{equation}
\frac{\partial E}{\partial\mathbf r_{ij}}=
\left.\frac{\partial\varepsilon^{(2)}(h^{(2)}_i)}{\partial\mathbf r_{ij}}\right|_{h^{(1)}}
+\left(\frac{\partial\varepsilon^{(1)}}{\partial h^{(1)}_i}+G_i\right)\cdot\frac{\partial h^{(1)}_i}{\partial\mathbf r_{ij}}.
\label{eq:chain}
\end{equation}
Partitioning the receivers into disjoint chunks $\mathcal C_1,\dots,\mathcal C_m$, exact energies, forces and virials
follow from three sweeps over the chunks (Algorithm~\ref{alg:threepass}). Pass A evaluates and stores
$H^{(1)}[\mathcal C]$ for each chunk from the edges incident on $\mathcal C$, without autograd. Pass B reads $H^{(1)}$ of
$\mathcal C$ and of its senders $\mathcal S(\mathcal C)=\bigcup_{i\in\mathcal C}\mathcal N(i)$, computes the layer-2
energy $e_2=\sum_{i\in\mathcal C}\varepsilon^{(2)}(h^{(2)}_i)$ and back-propagates it, accumulating the first term of
eq~\ref{eq:chain} into the pair forces and $\partial e_2/\partial H^{(1)}$ into $G$. Pass C recomputes $h^{(1)}_i$ for
$i\in\mathcal C$ with autograd and back-propagates the surrogate
$\ell=\sum_{i\in\mathcal C}\big[\varepsilon^{(1)}(h^{(1)}_i)+\langle G_i,h^{(1)}_i\rangle\big]$ with $G$ held fixed,
which yields the second term.

\begin{algorithm}[tbp]
\caption{Receiver-chunked three-pass MACE (exact energy and forces).}
\label{alg:threepass}
\begin{algorithmic}[1]
\For{each receiver chunk $\mathcal C$} \Comment{pass A, no autograd}
  \State $H^{(1)}[\mathcal C]\gets\Phi_1(\mathcal E(\mathcal C))$
\EndFor
\State $G\gets 0$
\For{each $\mathcal C$} \Comment{pass B}
  \State $e_2\gets\sum_{i\in\mathcal C}\varepsilon^{(2)}\big(\Phi_2(H^{(1)}[\mathcal C\cup\mathcal S(\mathcal C)],\mathcal E(\mathcal C))\big)$
  \State $\mathbf f_{ij}\mathrel{+}=-\partial e_2/\partial\mathbf r_{ij}$;\quad $G[\mathcal C\cup\mathcal S]\mathrel{+}=\partial e_2/\partial H^{(1)}$
\EndFor
\For{each $\mathcal C$} \Comment{pass C, recompute layer 1}
  \State $h\gets\Phi_1(\mathcal E(\mathcal C))$;\quad $\ell\gets\sum_{i\in\mathcal C}\varepsilon^{(1)}(h_i)+\langle G[\mathcal C],h\rangle$
  \State $\mathbf f_{ij}\mathrel{+}=-\partial\ell/\partial\mathbf r_{ij}$
\EndFor
\end{algorithmic}
\end{algorithm}

Wide edge intermediates exist only inside the active chunk, so the resident state consists of the coordinates, species,
an int32 neighbor list, $H^{(1)}$, $G$ and the forces. To avoid re-specialization of the cuEquivariance kernels and
recapture of CUDA graphs when neighbor counts fluctuate during MD, the node and edge arrays of a chunk are padded to
fixed capacities (25\% above the largest chunk) with zero-weight self-loops spread over distinct nodes. Coordinates and
edge vectors are kept in fp64 and cast to the model precision per chunk; without this, the fp32 force error grows with
box size.

\textbf{Other MACE models and $L$-layer networks.} The readout head of multi-head MACE models (e.g., MACE-MH-1) is
selected by name. For an $L$-layer message-passing network with node features $X_k$ ($k=1,\dots,L$), a forward sweep
evaluates $X_k[\mathcal C]$ from $X_{k-1}[\mathcal C\cup\mathcal S(\mathcal C)]$ layer by layer without autograd; the
backward sweep then recomputes layers $k=L,\dots,1$ chunk by chunk with autograd and the upstream term
$\langle G_k[\mathcal C],X_k[\mathcal C]\rangle$, accumulating pair forces and $G_{k-1}$. This requires $2L$ sweeps,
which reduce to three for $L=2$ because the last forward and the first backward sweep can be fused. We implemented this
schedule for the five-layer NequIP-type foundation model SevenNet-0\cite{park2024sevennet} with shape-padded chunk
tensors; its reference evaluation used fp64 edge vectors.

\subsection{LAMMPS Integration and Benchmarks}

The three-pass evaluator is coupled to LAMMPS through the Kokkos ML-IAP unified Python interface (LAMMPS 11 February
2026, commit 5c6749b; Kokkos 5.0.2; Figure~\ref{fig:overview}b). At every MD step the pair style filters the GPU
neighbor list by the model cutoff and calls the Python force routine. Between passes A, B and C, ghost values of
$H^{(1)}$ and ghost contributions to $G$ are exchanged with the forward and reverse communication of LAMMPS, split into
column blocks to respect 32-bit buffer indexing. Two changes to \texttt{mliap\_data\_kokkos.cpp} (Supplementary
Note~3), which avoid host mirrors of device-only per-pair arrays and grow the neighbor arrays with 10\% headroom instead
of reallocating at every increase of the pair count, reduce the LAMMPS-side overhead at $10^{6}$ atoms from 0.25 to
0.002~s per step. Two guards stop the run cleanly before an illegal memory access or a blow-up: before the evaluation
the neighbor-slot estimate $P((r_{\max}+s)/r_{\max})^{3}$, with $P$ pairs and skin $s$, must stay below
$0.95\cdot2^{31}$, and after it all forces must be finite and below 100~eV\,\AA$^{-1}$.

All benchmarks ran on single NVIDIA H200 GPUs (141~GB) with MACE 0.3.15, PyTorch 2.5.1 (CUDA 12.4), cuEquivariance
0.9.1 and Triton 3.1.0, and with GPUMD 5.8 for NEP89 (\texttt{nep89\_20250409}).\cite{fan2022gpumd,liang2026nep89}
Steady-state throughputs ($N_{\rm atoms}\times N_{\rm steps}/t_{\rm wall}$) exclude the first five steps (compilation
and CUDA-graph capture); those of NEP89 and of the students are means of three independent runs, whose spread is below
0.5\%. Production rates are the loop time reported by LAMMPS for whole segments of $10^{5}$ steps and include every
neighbor-list rebuild. Full software versions are given in the Supporting Information.

\subsection{Distillation, Supervision Protocol and Reference Calculations}

\textbf{Initial distillation.} The teacher (MACE-MPA-0 medium\cite{batatia2025foundation}) drives 1-fs Langevin MD of
small periodic cells (192--256 atoms) over a grid of temperatures and isotropic cell scalings: 260--420~K and length
factors 0.975--1.052 for water, 300--3500~K and 0.97--1.03 for SiO$_2$, and 300--1500~K and 0.98--1.02 for CuNi
(Supplementary Table~S5). Frames are sampled every 0.1~ps after 1~ps of equilibration and split into training and
held-out test sets; the water set is extended by teacher-labelled frames from NPT runs of earlier students. The students
are two-layer MACE models with $16\times0e+16\times1o$ hidden features, $\ell_{\max}=2$, correlation order 2, 8 Bessel
radial functions, a $[16,16]$ radial MLP, the Agnesi distance transform, ZBL pair repulsion and
$r_{\max}=4.0$--4.5~\AA. They are trained in fp32 for 120 epochs (batch size 4, learning rate 0.01, exponential moving
average with decay 0.99) with loss weights 1 (energy), 100 (forces) and $10^{4}$ (stress).

\textbf{On-the-fly supervision and cluster retraining.} The student runs NPT or NVT segments in LAMMPS; frames are
dumped at fixed intervals and recomputed by the student and by the exact teacher with the LAMMPS \texttt{rerun} command
(Figure~\ref{fig:overview}a). A segment triggers retraining if any of four criteria fails: (i) the system-averaged force
RMSE exceeds 4\% of the teacher's RMS force on the training frames (90~meV\,\AA$^{-1}$ for SiO$_2$) or
35~meV\,\AA$^{-1}$ for water; (ii) more than a fraction $f=10^{-5}$ of the atoms have
$|\Delta\mathbf F|=|\mathbf F_{\rm student}-\mathbf F_{\rm teacher}|>0.5$~eV\,\AA$^{-1}$; (iii) the virial pressure
deviates by more than 300--450~bar (3000~bar for SiO$_2$ at 3000~K); (iv) a safety stop fires (density below half its
initial value, temperature above three times the target, or non-finite or excessive forces). For frames of up to
$4\times10^{3}$ atoms the failed frames themselves enter the training set. For larger frames, up to four spheres of
radius $R_{\rm out}=13$~\AA, at least $2R_{\rm out}$ apart, are carved around the atoms with the largest errors (around
random atoms in the control arms) from each of the two worst frames. Every atom within 1.2 times the sum of covalent
radii of an included atom is added, up to three times, so that no bond is cut; the cluster (about 950 atoms for water
and 1,100 for SiO$_2$) is centred in a box with 8~\AA{} of vacuum and labelled with the teacher's energy and forces,
without stress. Per-atom force weights added to MACE 0.3.15 restrict the force loss to the atoms within
$R_{\rm in}=6$~\AA{} of the centre; cluster energies are kept out of the loss, and frames acquired on the fly are
weighted by 10. These settings apply to the SiO$_2$ experiments; the water hard-case runs used the all-atom force loss.
The student is fine-tuned from its current weights for 40--60 epochs at learning rate 0.005, and every segment starts
from the final configuration of the previous one.

\textbf{Controlled repair experiments.} In molten SiO$_2$ (3000~K, 1~bar, $r_{\max}=4.5$~\AA{} student) we ran five
2-ps NPT segments in 12,288-atom boxes with three seeds for each of three arms: a fixed student without retraining,
retraining on eight 13-\AA{} clusters per segment centred on the worst atoms, and the same with random centres. One
targeted run used $1.03\times10^{6}$ atoms. All models were evaluated on 33 independent frames of the teacher's own NPT
trajectories and on 36 DFT frames rich in coordination defects. In liquid water ($r_{\max}=4.25$~\AA{} student, 12,288
atoms) repair was tested in two states outside the training range: NVT at 450~K at ambient density and NVT at 300~K
compressed to 1.17~g\,cm$^{-3}$.

\textbf{Reference calculations.} DFT labels were computed with VASP 6.5.0 (GPU build), PBE, PAW PBE\_54 potentials and
a 520-eV cutoff, with settings compatible with the MPtrj data on which the teacher was trained (Supplementary
Table~S17). The classical comparisons used SPC/Fw (water), Vashishta and Tersoff (SiO$_2$) and EAM (CuNi) (Supplementary
Table~S18).

%% ====================================================================================================================
\section{Results and Discussion}

We first validate the exactness, memory scaling and label fidelity of the three-pass teacher
(section~\ref{sec:exact}), then show what whole-system checks uncover in molten SiO$_2$ (section~\ref{sec:find}),
evaluate on-the-fly repair in controlled arms and at two system sizes (section~\ref{sec:repair}), analyze the
trade-off between check interval, detection delay and cost (section~\ref{sec:cost}), and finally benchmark the
students and the generality of the chunked evaluation (sections~\ref{sec:speed} and \ref{sec:general}).

\subsection{Exactness, Memory Scaling and Cluster Label Fidelity}
\label{sec:exact}

\begin{figure*}[tbp]
\centering
\includegraphics[width=\linewidth]{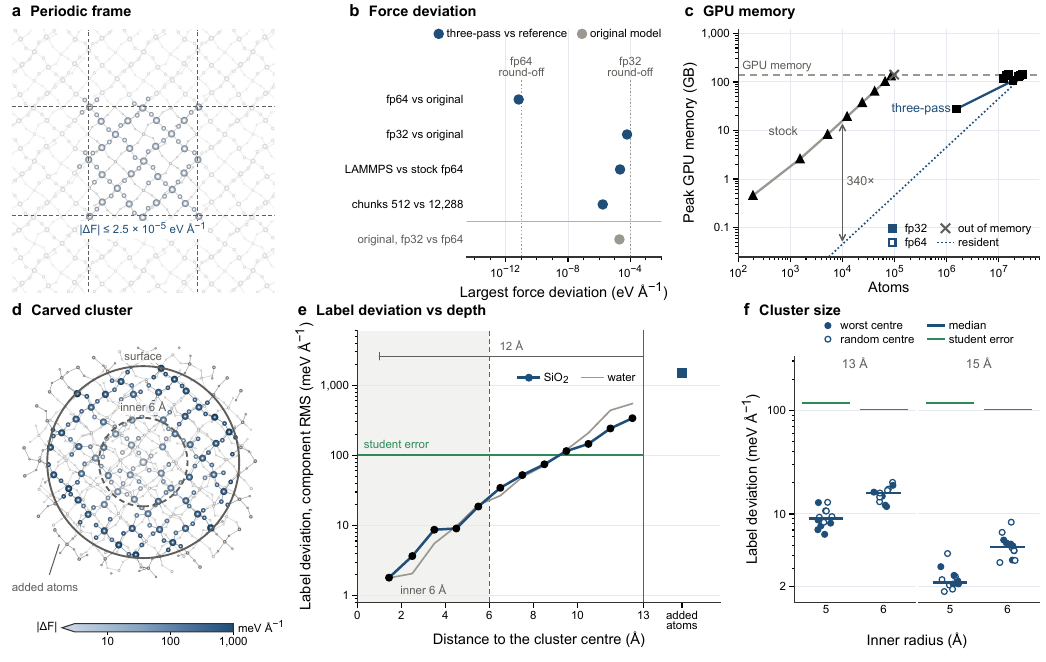}
\caption{Exactness, memory scaling and cluster label fidelity of the three-pass teacher. (a) A periodic frame has no
surface: the 192-atom SiO$_2$ cell of the exactness test with its periodic images. The largest force deviation of the
production evaluation (fp32, 512-knot tables) from the original fp64 model over all atoms is
$\le2.5\times10^{-5}$~eV\,\AA$^{-1}$, a bound rather than a per-atom map. (b) Largest force deviation against the
original model in fp64 and fp32, inside LAMMPS against the stock fp64 model, and for different chunk sizes, for
replicated water boxes of up to $1.53\times10^{7}$ atoms; dotted lines, round-off references. (c) Peak GPU memory
against $N$ for water on one H200: stock MACE-MPA-0 through the LAMMPS ML-IAP interface (grey; largest run 86,016 atoms,
out of memory at 98,304) and the three-pass evaluation (navy; filled, fp32; open, fp64; dotted, its resident part, 340
times less per added atom). (d) A carved 13-\AA{} SiO$_2$ cluster (1,104 atoms, centred on a worst atom) coloured by
$|\mathbf F_{\rm cluster}-\mathbf F_{\rm periodic}|$: pale in the interior, dark at the surface; dashed circle, the
6-\AA{} loss region. (e) Component RMS label deviation against distance from the cluster centre for SiO$_2$ (navy) and
water (grey) clusters, with the initial student's error in the inner 6~\AA{} (green line) and the teacher's 12-\AA{}
receptive field measured from the surface (bracket). (f) Inner-region label deviation of SiO$_2$ clusters with outer
radii of 13 and 15~\AA{} (filled, worst-atom centres; open, random centres; bars, medians).}
\label{fig:exact}
\end{figure*}

\textbf{Exactness.} We verify the three-pass evaluation atom by atom on periodically replicated water boxes of up to
$1.53\times10^{7}$ atoms, in which every atom must experience the force of its base-cell image
(Figure~\ref{fig:exact}a,b; Supplementary Table~S1). Against the original full-graph MACE-MPA-0 model the largest force
deviation is $6.7\times10^{-12}$~eV\,\AA$^{-1}$ in fp64 and $6.2\times10^{-5}$~eV\,\AA$^{-1}$ in fp32, at the level of
the original model's own fp32--fp64 difference, and it neither grows with $N$ nor depends on the chunk size. Inside
LAMMPS, the production path (fp32 with radial weights tabulated on a 512-knot cubic Hermite grid) agrees with the stock
fp64 ML-IAP model within $2.2\times10^{-5}$~eV\,\AA$^{-1}$, with comparable contributions from fp32 rounding and the
tables. Central finite differences confirm that forces and virial are derivatives of the computed energy (Supplementary
Tables~S2 and S3). NVE dynamics of water at 1~fs are stable over 100~ps, and NPT water gives $0.9140\pm0.0004$ against
$0.9151\pm0.0005$~g\,cm$^{-3}$ for the original model, with matching radial distribution functions (Supplementary
Figure~S1).

\textbf{Memory and capacity.} Because per-edge intermediates are discarded after each chunk, resident memory grows
linearly with $N$ at 4.57~kB per atom in fp32 (8.67~kB in fp64), plus a working buffer for the active chunk
(Figure~\ref{fig:exact}c). The stock cuEquivariance ML-IAP path needs 1.55~MB per atom and fits 86,016 atoms for a
3-step evaluation and 41,472 atoms for an MD run; the three-pass schedule reduces the marginal memory per added atom
340-fold (310-fold including working buffers at peak). On one H200 the standalone evaluation reaches
$2.86\times10^{7}$ atoms in fp32 and $1.53\times10^{7}$ in fp64, 1.5 times as many atoms as NEP89
($\approx1.86\times10^{7}$). Inside LAMMPS the limit is $8.96\times10^{6}$ atoms, set by 32-bit indexing of the Kokkos
neighbor list rather than by GPU memory; two H200 GPUs of one node run $1.57\times10^{7}$ atoms at
$2.65\times10^{5}$~atom-steps\,s$^{-1}$, 1.8--1.9 times the rate of one GPU (Supplementary Note~1).

\textbf{Fidelity of carved cluster labels.} When a segment fails, clusters carved around the failing environments are
labelled by the teacher in vacuum. Their labels are reliable only if the outer shell screens the loss region from the
surface over the teacher's two-layer receptive field ($2\times6$~\AA{} $=12$~\AA; Figure~\ref{fig:exact}d,e). For
metallic CuNi a 7-\AA{} cluster suffices (2~meV\,\AA$^{-1}$ deviation of the centre atom from the periodic frame). For
water and SiO$_2$ it does not: at 7~\AA{} the cluster signal deviates by 42 and 119~meV\,\AA$^{-1}$ from the bulk
signals of 90 and 41~meV\,\AA$^{-1}$ (Supplementary Table~S7). With $R_{\rm out}=13$~\AA{} and the force loss restricted
to $R_{\rm in}=6$~\AA, the label deviation over all inner atoms of the SiO$_2$ clusters falls to 16~meV\,\AA$^{-1}$,
one sixth of the initial student's error on the same atoms (102~meV\,\AA$^{-1}$; Figure~\ref{fig:exact}f,
Supplementary Table~S8). 15-\AA{} clusters lower this residual to 5~meV\,\AA$^{-1}$ but give the same retrained student
on independent frames (61.8 against 61.3~meV\,\AA$^{-1}$; Supplementary Note~4), so 13-\AA{} clusters with a 6-\AA{}
loss region capture the bulk correction.

\subsection{Whole-System Checks Uncover Persistent and Isolated Local Failures}
\label{sec:find}

\begin{figure*}[tbp]
\centering
\includegraphics[width=\linewidth]{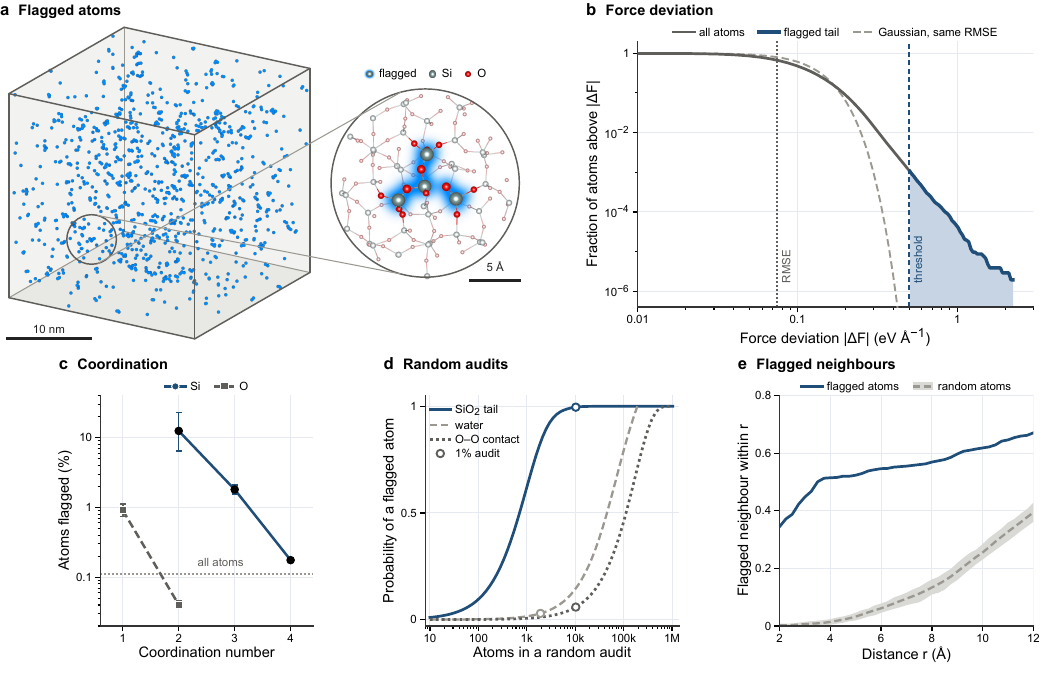}
\caption{Whole-system checks locate sparse, clustered failure tails that averaged metrics and random audits miss.
$1.03\times10^{6}$-atom molten SiO$_2$ frame (3000~K, 2~ps) driven by the initial $r_{\max}=4.5$~\AA{} student.
(a) The 1,140 flagged atoms ($|\Delta\mathbf F|>0.5$~eV\,\AA$^{-1}$, blue) in the 25-nm periodic box, with one flagged
site of seven under-coordinated atoms enlarged on the right. (b) Fraction of atoms whose $|\Delta\mathbf F|$ exceeds a
given value, with the force RMSE of the frame (74~meV\,\AA$^{-1}$, dotted), the 0.5-eV\,\AA$^{-1}$ threshold (dashed)
and an isotropic Gaussian error of the same RMSE (grey). (c) Share of flagged atoms in each coordination class of Si
and O (error bars, Wilson 95\% intervals; dotted line, overall flagged fraction 0.11\%). (d) Probability $P_{\rm hit}$
that a random audit of $m$ atoms contains at least one flagged atom, for the SiO$_2$ tail, the ambient-water tail and
the isolated six-atom O--O contact (open circles, 1\% audits). (e) Fraction of flagged atoms with another flagged atom
within distance $r$ (navy), compared with 100 random sets of 1,140 atoms (grey; dashed, mean; band, 95\% range).}
\label{fig:find}
\end{figure*}

In a $1.03\times10^{6}$-atom NPT simulation of molten SiO$_2$ at 3000~K driven by the initial $r_{\max}=4.5$~\AA{}
student, the system-averaged force RMSE against the teacher stays at 66--74~meV\,\AA$^{-1}$, below the
90-meV\,\AA$^{-1}$ threshold. The whole-system check nevertheless flags a heavy tail: after 2~ps, 1,140 atoms
($1.1\times10^{-3}$ of the system) exceed $|\Delta\mathbf F|=0.5$~eV\,\AA$^{-1}$ (up to 2.3~eV\,\AA$^{-1}$), two orders
of magnitude above the $10^{-5}$ criterion, whereas an isotropic Gaussian error with the same RMSE would place only
$7\times10^{-4}$ atoms of the whole frame above the threshold (Figure~\ref{fig:find}a,b).

The flagged atoms are under-coordinated and clustered in space (Figure~\ref{fig:find}c,e). Si with fewer than four O
within 2.0~\AA{} and O with fewer than two Si make up 2.9\% and 1.5\% of all atoms but 24\% and 25\% of the flagged
ones, an 8- and 17-fold enrichment. 51\% of the flagged atoms have another flagged atom within 4~\AA, against 1.4\% for
random sets of as many atoms, and together they form 771 sites. A 12,288-atom box shows the same signature, and the tail
is persistent: flagged atoms are present in 197 of 201 frames over 20~ps.

Comparing whole-system checks with uniform random audits of $m$ atoms separates detection from localization
(Figure~\ref{fig:find}d, Supplementary Table~S9). For a frame with $n$ flagged atoms out of $N$, an audit contains at
least one flagged atom with probability $P_{\rm hit}=1-\binom{N-n}{m}/\binom{N}{m}\approx1-e^{-mn/N}$. For a diffuse
tail of fixed fraction $n/N\approx10^{-3}$, an audit of $m=10^{3}$ atoms detects the failure in about 60\% of the
frames at any box size but contains on average only 1.1 of the 1,140 flagged atoms of the $10^{6}$-atom frame. For an
isolated failure, such as an unphysical O--O contact in the million-atom run where only $n=6$ atoms deviate by more than
10~eV\,\AA$^{-1}$, even an audit of 1\% of the box ($10^{4}$ atoms) misses the failure 94\% of the time
($P_{\rm hit}=6\%$). Only a whole-system check locates every failing site.

\subsection{On-the-Fly Retraining Repairs Failures up to the Teacher's Accuracy}
\label{sec:repair}

\begin{figure*}[tbp]
\centering
\includegraphics[width=\linewidth]{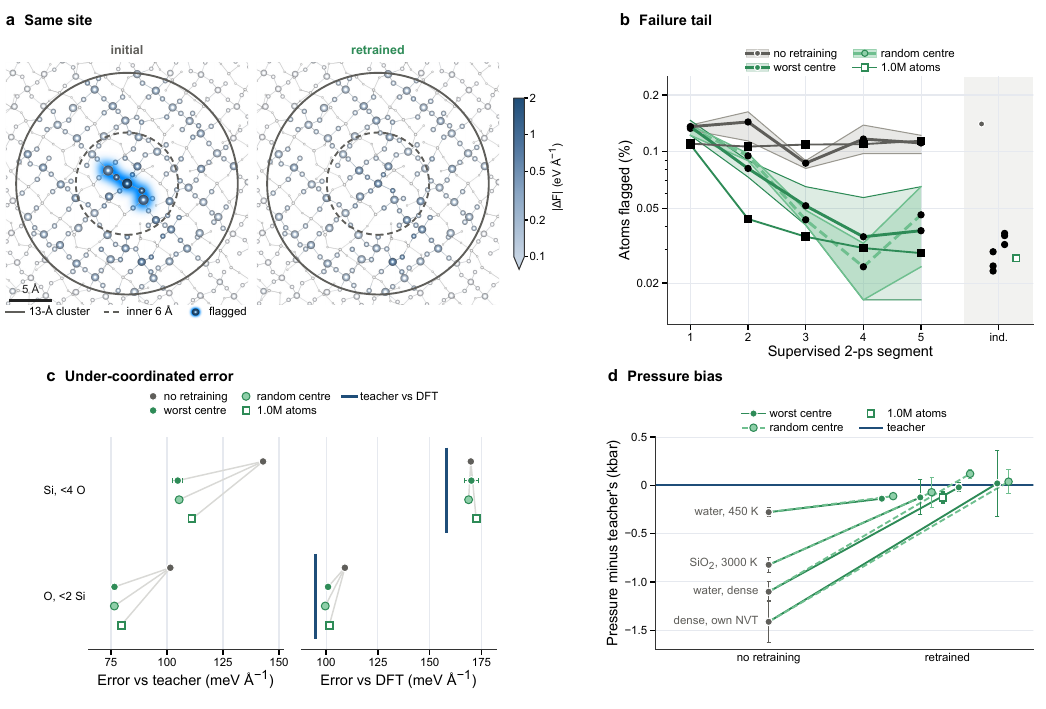}
\caption{On-the-fly cluster retraining repairs local failures and pressure biases up to the teacher's accuracy.
Molten SiO$_2$ at 3000~K without retraining (grey), with clusters centred on the worst atoms (dark green) and on random
atoms (light green), in 12,288-atom boxes (three seeds) and a $1.03\times10^{6}$-atom box (open squares). (a) Force
deviation around one failure site of an independent teacher frame for the initial (left) and the retrained student
(right); solid circle, 13-\AA{} cluster; dashed circle, 6-\AA{} loss region. (b) Flagged fraction
($|\Delta\mathbf F|>0.5$~eV\,\AA$^{-1}$) over five 2-ps segments and on the 33 independent teacher frames (ind.); bands
span the three seeds. (c) Component force RMSE on under-coordinated Si and O against the teacher (left) and against DFT
(right; navy bar, the teacher's own error). (d) Pressure bias relative to the teacher on identical frames (SiO$_2$,
water at 450~K, compressed water at 1.17~g\,cm$^{-3}$) and in paired 20-ps NVT runs (dense, own NVT); points, means;
error bars, standard deviations over frames or runs.}
\label{fig:repair}
\end{figure*}

\textbf{Repair of diffuse tails and pressure biases.} Fine-tuning on eight teacher-labelled 13-\AA{} clusters after
each 2-ps segment repairs the SiO$_2$ failure in all three seeds of the 12,288-atom runs and in the
$1.03\times10^{6}$-atom run (Figure~\ref{fig:repair}a,b; Supplementary Tables~S10 and S14). On the 33 independent
teacher frames, retraining lowers the force RMSE from 74.9 to 60.2--61.2~meV\,\AA$^{-1}$ in the 12,288-atom runs (by a
fifth) and to 66.5~meV\,\AA$^{-1}$ in the $10^{6}$-atom run, shrinks the flagged fraction by three quarters (from
$1.4\times10^{-3}$ to $2.6$--$3.5\times10^{-4}$), lowers the error on under-coordinated Si from 143 to
105~meV\,\AA$^{-1}$ (Figure~\ref{fig:repair}c), and reduces the pressure bias on identical configurations from
$-823\pm77$~bar to about $-0.1$~kbar (Figure~\ref{fig:repair}d).

The repair is equally clear in the two water states (Figure~\ref{fig:repair}d, Supplementary Figure~S10). In compressed
water the force RMSE on independent teacher frames falls from 37.1 to 25.8--26.1~meV\,\AA$^{-1}$ and a static pressure
bias of $-1.10$~kbar is removed. The repair carries over to the student's own dynamics: in three paired 20-ps NVT runs
at 1.17~g\,cm$^{-3}$ the pressure of the initial student lies $1.41\pm0.13$~kbar below the teacher's, whereas the
retrained students deviate by $+0.02\pm0.20$~kbar (worst-atom clusters) and $+0.04\pm0.07$~kbar (random clusters; mean
and standard error over the runs; Supplementary Table~S12).

\textbf{Three qualifications.} The controlled experiments establish three qualifications.
(1) \textit{Diffuse tails do not need targeted placement.} For the distributed tails of SiO$_2$ and compressed water,
clusters around random atoms work as well as clusters around the worst atoms (60.2$\pm$0.1 and 61.2$\pm$0.4~meV\,\AA$^{-1}$
in SiO$_2$; Figure~\ref{fig:repair}b--d). A 13-\AA{} cluster holds 950--1,150 atoms (the inner regions of eight
clusters about 470 in all), so randomly placed clusters sample the rare environments anyway; the improvement comes from
triggering retraining and from labels that reproduce the bulk.
(2) \textit{Isolated failures need the worst atoms.} In the third segment of the million-atom run two pairs of singly
coordinated O atoms approached to 0.82~\AA. The initial student repels the same configuration only weakly
(13--20~eV\,\AA$^{-1}$), and two retraining rounds whose clusters sampled other environments turned this into an
attraction ($-5$ to $-7$~eV\,\AA$^{-1}$, against a repulsion of 171--178~eV\,\AA$^{-1}$ by the teacher; Supplementary
Movie~S2, Supplementary Figure~S8). Eight randomly placed clusters would bring one of the six contact atoms into the
loss with a probability below 0.3\% per round. Clusters centred on the worst atoms captured the contact in the third
and fourth rounds, after which the student repels it as strongly as the teacher (167--168~eV\,\AA$^{-1}$ at 0.82~\AA);
on 28 untrained configurations with O--O distances of 0.85--1.45~\AA{} the correction holds in sign and magnitude but
not in detail (Supplementary Table~S11, Supplementary Figure~S9).
(3) \textit{Repair converges to the teacher, not to DFT.} On the 36 DFT frames, retraining improves the student on
under-coordinated O from 109 to 100~meV\,\AA$^{-1}$, towards the teacher's 95, but leaves it at 166--174~meV\,\AA$^{-1}$
on under-coordinated Si, where the teacher itself errs by 158~meV\,\AA$^{-1}$ (Figure~\ref{fig:repair}c). Supervision
repairs distillation gaps; correcting errors of the foundation model requires fine-tuning the teacher on DFT data.

\subsection{Check Interval, Detection Delay and Overhead}
\label{sec:cost}

\begin{figure*}[tbp]
\centering
\includegraphics[width=\linewidth]{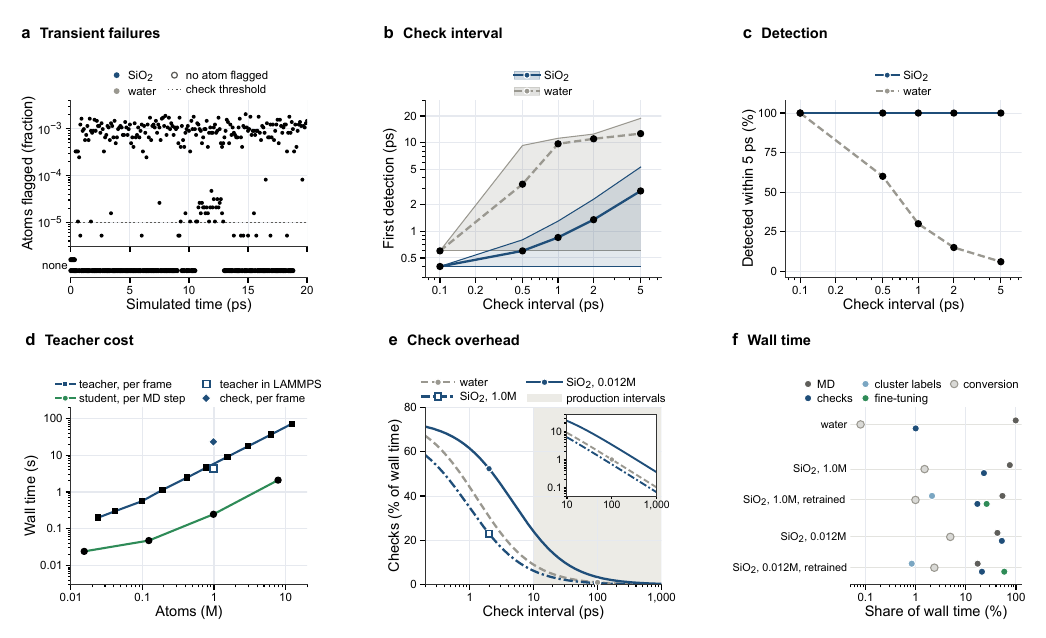}
\caption{Check interval, detection delay and cost. (a) Flagged fraction on frames spaced by 0.1~ps without
retraining: persistent in SiO$_2$ at 3000~K (12,288 atoms, navy), transient in water at 300~K (192,000 atoms, grey);
dotted line, the $10^{-5}$ threshold. (b) First failed check and (c) share of check offsets that detect the failure
within 5~ps, against the check interval (line, median over offsets; band, range). (d) Wall time against system size on
one H200 for one standalone teacher evaluation (navy squares), the teacher inside LAMMPS (open square), one student MD
step (green circles) and a complete production check including start-up and I/O (diamond). (e) Share of wall time spent
on checks against the check interval (curves, cost model; markers, measured runs; shading and inset, production
intervals of 10~ps and more). (f) Wall time by stage (MD, checks, cluster labels, fine-tuning, model conversion) for the
1-ns water production run and the SiO$_2$ supervision runs.}
\label{fig:cost}
\end{figure*}

The SiO$_2$ tail is persistent and is detected at any check interval, but the failures in ambient water (300~K, 192,000
atoms) are transient: only 44 of 189 frames spaced by 0.1~ps contain atoms with $|\Delta\mathbf F|>0.5$~eV\,\AA$^{-1}$,
at most nine per frame (Figure~\ref{fig:cost}a). Sparse checks therefore see transient tails late: checks every 0.5, 1
and 2~ps detect the failure after a median of 3.4, 9.7 and 11~ps, and at 5-ps intervals 42\% of the check offsets never
detect it (Figure~\ref{fig:cost}b,c). Passed sparse checks thus rule out persistent failures rather than transient
spikes; criteria based on the persistence of flagged atoms over consecutive checks or on their absolute number would
separate the two cases.

Whole-system checks are inexpensive at scale (Figure~\ref{fig:cost}d--f). One exact teacher evaluation of $10^{6}$
atoms takes 4.4~s inside LAMMPS, as long as 18 MD steps of the student (0.25~s per step), and a complete check including
LAMMPS start-up and file I/O takes 23~s. In the 1-ns production run of $9.83\times10^{5}$ water atoms checked every
100~ps, supervision costs 1.3\% of the 64.4-h wall time, of which the checks themselves take 1.0\%
(Figure~\ref{fig:cost}e,f). When retraining is triggered in every 2-ps segment of the $1.03\times10^{6}$-atom SiO$_2$
run, MD takes 54\% of the wall time, checks 17\%, cluster labelling 2.1\%, fine-tuning 26\% and model conversion 1\%
(Supplementary Table~S14, Supplementary Figure~S11).

\subsection{Speed and DFT Accuracy of the Distilled Students}
\label{sec:speed}

Multiple time stepping\cite{tuckerman1992respa} with a cheap inner potential does not accelerate MACE-MPA-0 effectively
in water: the force difference to NEP89 is 0.49~eV\,\AA$^{-1}$ (RMS), and with MACE-MP-0b2-small the integration is
stable only for outer-to-inner step ratios of 2--3. Distilled short-range MACE students avoid this limitation. A NEP
student trained on the same water labels with GPUMD's default recipe ran eight times faster than the MACE student but
reached 51~meV\,\AA$^{-1}$ against the teacher and did not reproduce liquid water in 300-K dynamics (Supplementary
Note~4), whereas the MACE students were stable from the first attempt. The NEP recipe was not tuned.

\begin{figure*}[tbp]
\centering
\includegraphics[width=\linewidth]{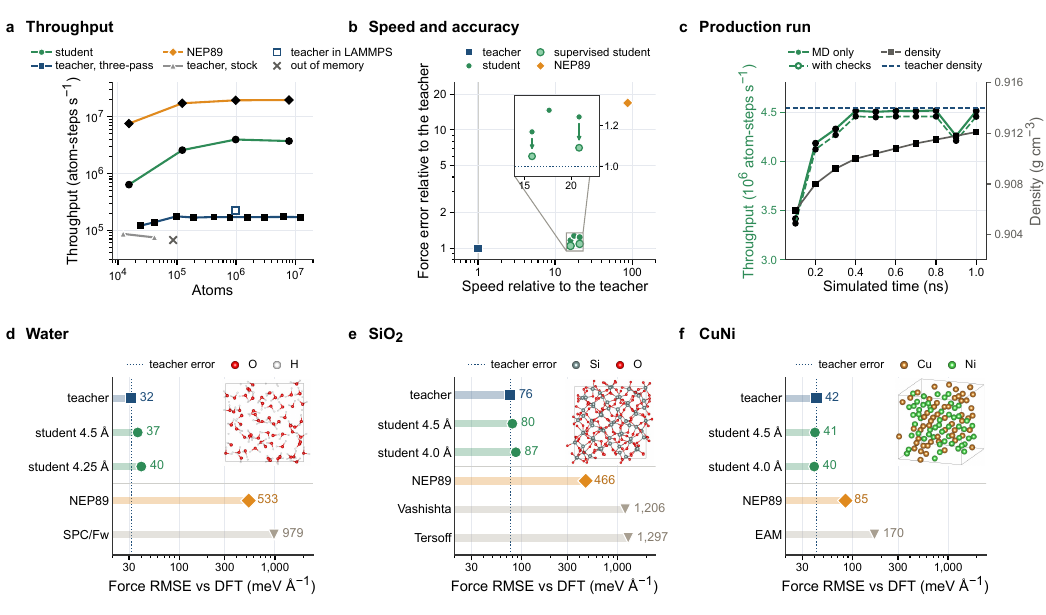}
\caption{Throughput and DFT accuracy of the distilled MACE students. (a) MD throughput against system size on one
H200: water student ($r_{\max}=4.25$~\AA), NEP89, the three-pass teacher (standalone; open square, inside LAMMPS) and
stock MACE-MPA-0 (out of memory at 86,016 atoms, cross). (b) Speed and force error against DFT (water, 20 frames),
both relative to the teacher, for students with $r_{\max}=4.5$, 4.25 and 4.0~\AA{} before (dark green) and after
supervision (light green) and for NEP89 (inset, linear scale). (c) Throughput (MD alone and with checks) and NPT
density (teacher, dashed) over the 1-ns supervised run of $9.83\times10^{5}$ water atoms. (d--f) Component force RMSE
against DFT on held-out periodic frames for (d) water (20 frames of 192 atoms), (e) molten SiO$_2$ (20 frames of 192
atoms) and (f) CuNi (4 frames of 108 atoms): teacher (dotted), students, NEP89 and classical force fields (SPC/Fw,
Vashishta, Tersoff, EAM); insets, the test cells.}
\label{fig:speed}
\end{figure*}

Because the cost of passes B and C scales with the number of edges, the student cutoff sets the trade-off between speed
and accuracy (Figure~\ref{fig:speed}a,b; Table~\ref{tab:summary}). On one H200 at $\sim10^{6}$ atoms, the water students
run 16--21 times faster than the exact teacher, and across the three systems the students run four to nine times slower
than NEP89 with 0.95--1.28 times the teacher's force error against DFT; on water they are 13--15 times more
accurate than NEP89 (Figure~\ref{fig:speed}d--f, Supplementary Figure~S12).

\begin{table}[tbp]
\caption{Performance and Accuracy of the Exact Teacher, the Distilled Students, NEP89 and Classical Force Fields on One
H200 GPU$^{a}$}
\label{tab:summary}
\scriptsize\setlength{\tabcolsep}{3.5pt}
\begin{tabular}{@{}llccccccc@{}}
\toprule
system & model & $r_{\max}$ & throughput & vs NEP89 & $F$ vs & $F$ vs & $E$ vs & $\rho$ \\
 & & (\AA) & (atom-steps\,s$^{-1}$) & & teacher & DFT & DFT & (g\,cm$^{-3}$) \\
\midrule
water & teacher (MACE-MPA-0) & 6.0 & $2.25\times10^{5}$ & 1/87 & -- & 31.6 & 0.65 & 0.914 \\
 & student & 4.5 & $3.56\times10^{6}$ & 1/5.5 & 25.7 & 36.9 (33.1$^{b}$) & 0.62 & $-1.2\%$$^{c}$ \\
 & student & 4.25 & $3.97\times10^{6}$ & 1/4.9 & 29.7 & 40.3 & 0.71 & $-1.6\%$$^{c}$ \\
 & student & 4.0 & $4.69\times10^{6}$ & 1/4.2 & 28.7 & 34.5$^{b}$ & -- & $-2.1\%$$^{c}$ \\
 & NEP89 & -- & $1.96\times10^{7}$ & 1 & -- & 532.7 & 4.3 & -- \\
 & SPC/Fw & -- & -- & -- & -- & 978.5 & 3.8 & -- \\
\midrule
SiO$_2$ & teacher (MACE-MPA-0) & 6.0 & -- & -- & -- & 76.1 & 2.16 & 2.0823 \\
 & student & 4.5 & $4.55\times10^{6}$ & 1/6.9 & 74.2 & 80.0 & 2.16 & 2.0827 \\
 & student & 4.0 & $5.81\times10^{6}$ & 1/5.4 & 75.3 & 86.9 & 2.07 & 2.0815 \\
 & NEP89 & -- & $3.12\times10^{7}$ & 1 & -- & 466.1 & 30 & -- \\
 & Vashishta / Tersoff & -- & -- & -- & -- & 1205.6 / 1297.0 & 79 / 56 & -- \\
\midrule
CuNi & teacher (MACE-MPA-0) & 6.0 & -- & -- & -- & 42.3 & 1.55 & -- \\
 & student & 4.5 & $3.03\times10^{6}$ & 1/8.8 & 22.5 & 40.7 & 1.21 & 8.4765 \\
 & student & 4.0 & $5.32\times10^{6}$ & 1/5.0 & 20.2 & 40.0 & 1.38 & 8.4805 \\
 & NEP89 & -- & $2.68\times10^{7}$ & 1 & -- & 84.9 & 3.6 & -- \\
 & EAM & -- & -- & -- & -- & 170.1 & 2.0 & -- \\
\bottomrule
\end{tabular}

\smallskip
{\footnotesize $^{a}$Throughputs at $9.83\times10^{5}$ (water, 300~K), $1.03\times10^{6}$ (SiO$_2$, 3000~K) and
$1.00\times10^{6}$ atoms (CuNi, 1000~K); means of three runs for the students and NEP89 (spread below 0.5\%). $F$:
component force RMSE (meV\,\AA$^{-1}$) on held-out periodic frames; $E$: energy RMSE against DFT with the mean offset
removed (meV\,atom$^{-1}$). $\rho$: NPT density of a 1,536-atom box. --: not measured or not applicable. The stock
MACE-MPA-0 evaluation has the same errors as the three-pass teacher; in LAMMPS it reaches $6.5\times10^{4}$
atom-steps\,s$^{-1}$ and fits at most 41,472 atoms in MD. $^{b}$After supervision. $^{c}$Relative to the teacher's
density in NPT runs of 30--50~ps.}
\end{table}

In the 1-ns supervised NPT production run of $9.83\times10^{5}$ water atoms ($r_{\max}=4.25$~\AA{} student, ten
segments of 100~ps), the steady segments ran at $4.50$--$4.52\times10^{6}$~atom-steps\,s$^{-1}$
($4.45$--$4.46\times10^{6}$ including model builds and checks), and the whole run, including the equilibration segment,
at $4.24\times10^{6}$~atom-steps\,s$^{-1}$, or 0.37~ns per day (Figure~\ref{fig:speed}c, Supplementary Movie~S3). All
ten checks passed without retraining: the force RMSE on the worst checked frame fell from 28.7 to 25.8~meV\,\AA$^{-1}$
as the box relaxed, and at most three atoms in a million exceeded 0.5~eV\,\AA$^{-1}$. In water, shortening the cutoff
from 4.5 to 4.0~\AA{} lowers the NPT density by 1.2--2.1\% relative to the teacher, consistent with the interactions
between 4 and 6~\AA{} that the shorter cutoffs omit. In SiO$_2$ and CuNi the 4.0-\AA{} students keep the teacher's
density within 0.04\% and run at $5.81$ and $5.32\times10^{6}$~atom-steps\,s$^{-1}$ (1/5.4 and 1/5.0 of NEP89).

\subsection{Arithmetic Ceiling, Other Architectures and Engines}
\label{sec:general}

\textbf{Arithmetic ceiling of the exact teacher.} Counting multiply--adds in the cuEquivariance descriptors of
MACE-MPA-0 on water gives $\approx293$~kFLOP per edge and $\approx19$~MFLOP per atom per pass, $\approx45$~MFLOP per
atom-step for the three passes, of which the recomputation of layer 1 in pass C adds about 10\% (Supplementary Note~2,
Supplementary Table~S4). At the 67-TFLOP\,s$^{-1}$ fp32 peak of the H200 this caps the exact model at
$1.5\times10^{6}$~atom-steps\,s$^{-1}$ ($0.6$--$1.1\times10^{6}$ at 40--60\% kernel utilization), an order of magnitude
below NEP89 ($2.0\times10^{7}$). Four exact optimizations---a species-block gather in place of the dense 89-element
one-hot self-connection product, node operations restricted to the receivers of the chunk, radial weights tabulated on a
512-knot cubic Hermite grid and evaluated by a Triton kernel, and fixed-shape chunk passes captured in CUDA
graphs---bring the evaluation in LAMMPS to $2.0$--$2.3\times10^{5}$~atom-steps\,s$^{-1}$, 11--15\% of this ceiling.

\textbf{Other MACE models and NequIP-type networks.} The chunked evaluator is model-agnostic
(Table~\ref{tab:models}). Without code changes it runs MACE-MATPES-r$^2$SCAN-0, MACE-MATPES-PBE-0\cite{kaplan2025matpes}
and the four-times-wider multi-head MACE-MH-1\cite{batatia2025mh1} on $9.83\times10^{5}$ water atoms in LAMMPS and
reproduces the full-graph fp64 forces to $4.0$--$4.4\times10^{-13}$~eV\,\AA$^{-1}$. Applied to the five-layer
SevenNet-0,\cite{park2024sevennet} the $2L$-pass schedule reduces the resident memory from 5.7~MB per atom (e3nn
backend; out of memory at 24,576 atoms) to 8.7~kB per atom in fp32. It evaluates $3.1\times10^{6}$ atoms standalone,
31 times as many as the original with its best backend, and runs $9.83\times10^{5}$ atoms in LAMMPS at
$8.5\times10^{4}$~atom-steps\,s$^{-1}$ with a 23-GB peak, matching the speed of SevenNet's own interface with ten times
its capacity.

\begin{table}[tbp]
\caption{Other Foundation Models and Architectures through the Same Exact Evaluation (Water, One H200)$^{a}$}
\label{tab:models}
\footnotesize
\begin{tabular}{@{}lcccccc@{}}
\toprule
model & layers & replica check & LAMMPS & resident & throughput & peak \\
 & & (eV\,\AA$^{-1}$) & (eV\,\AA$^{-1}$) & (kB/atom) & (atom-steps\,s$^{-1}$) & (GB) \\
\midrule
MACE-MPA-0 & 2 & $6.7\times10^{-12}$ $^{b}$ & $2.1\times10^{-5}$ & 4.6 & $2.2\times10^{5}$ & 136$^{c}$ \\
MACE-MATPES-r$^2$SCAN-0 & 2 & $4.2\times10^{-13}$ & $1.7\times10^{-5}$ & 4.6 & $1.97\times10^{5}$ & 31 \\
MACE-MATPES-PBE-0 & 2 & $4.0\times10^{-13}$ & $1.6\times10^{-5}$ & 4.6 & $1.97\times10^{5}$ & 31 \\
MACE-MH-1 (OMat24 head) & 2 & $4.4\times10^{-13}$ & $2.2\times10^{-5}$ & 16.9 & $1.5\times10^{4}$ & 52 \\
SevenNet-0 (NequIP-type) & 5 & $4.4\times10^{-13}$ & $4.7\times10^{-6}$ & 8.7 & $8.5\times10^{4}$ $^{d}$ & 23 \\
\bottomrule
\end{tabular}

\smallskip
{\footnotesize $^{a}$Replica check: largest force deviation over all 12,288 atoms of a $4\times4\times4$ replica of the
192-atom cell from the original full-graph model on the base cell, fp64. LAMMPS: largest force deviation of the
three-pass model in LAMMPS (fp32) from the fp64 reference, 192 atoms. Resident: per-atom memory kept for all atoms,
fp32. Throughput and peak memory at 983,040 atoms in LAMMPS. $^{b}$All $1.53\times10^{7}$ atoms of the replica.
$^{c}$Chunks of 65,536 receivers; 81~GB at $1.76\times10^{5}$~atom-steps\,s$^{-1}$ with chunks of 32,768
(Supplementary Table~S15). $^{d}$SevenNet's cuEquivariance backend with padded chunk shapes;
$1.3\times10^{4}$~atom-steps\,s$^{-1}$ and 39~GB with its e3nn backend.}
\end{table}

\textbf{Comparison with a concurrent compiled engine.} We ran our evaluator and Symmetrix-XL\cite{fu2026symmetrixxl}
back to back on the same H200 and water boxes (Supplementary Note~5, Supplementary Table~S15). At $9.83\times10^{5}$
atoms Symmetrix-XL runs at $2.10\times10^{5}$~atom-steps\,s$^{-1}$ with 77~GB, and ours at $2.02\times10^{5}$ with
136~GB (chunks of 65,536) or $1.76\times10^{5}$ with 81~GB (chunks of 32,768); both reach $\sim10^{7}$ atoms in LAMMPS
and are 3.3--3.6 times faster than the stock ML-IAP path. Symmetrix-XL evaluates the model through 256-point radial
splines and compact edge geometry, so its forces deviate from the fp64 reference by up to $1.6\times10^{-4}$ (fp32) and
$6\times10^{-5}$~eV\,\AA$^{-1}$ (fp64), against $2\times10^{-5}$ and $10^{-14}$~eV\,\AA$^{-1}$ for ours. Our
PyTorch-based chunked evaluation also extends to multi-layer NequIP-type networks and carries the exact supervision and
cluster-retraining pipeline of this work.

%% ====================================================================================================================
\section{Conclusions}

Supervising every atom of a large-scale MD simulation with an exact equivariant foundation model finds and repairs
local failures of fast distilled potentials at small cost: 1.3\% of the run time at one check per 100~ps for $10^{6}$
atoms. This required removing the memory limit of message-passing potentials. The receiver-chunked three-pass schedule
discards resident edge intermediates without changing the model, reduces the marginal memory per atom 340-fold to
4.57~kB, and scales MACE-MPA-0 to $2.86\times10^{7}$ atoms standalone and $8.96\times10^{6}$ atoms in LAMMPS on one
H200. Exact two-layer MACE inference still needs $\approx45$~MFLOP per atom-step, which places its arithmetic ceiling
an order of magnitude below descriptor-based potentials; distilled short-range MACE students narrow this gap to four to
nine times the cost of NEP89 at close to the teacher's accuracy against DFT.

The controlled experiments establish three principles for on-the-fly supervision and repair. (1) The system-averaged
RMSE hides sparse failure tails ($\sim10^{-3}$ of the atoms in molten SiO$_2$); a 1\% random audit detects such a
diffuse tail in a large box but misses an isolated failure, the O--O contact of six atoms in $10^{6}$, 94\% of the
time, whereas a whole-system check locates every failing site. (2) Carved clusters must screen the loss region over
the teacher's receptive field: 13-\AA{} clusters with the force loss restricted to the inner 6~\AA{} reproduce the bulk
labels, remove three quarters of the SiO$_2$ failure tail and remove the kilobar-level pressure biases in SiO$_2$ and
compressed water. Random placement suffices for diffuse tails, whereas isolated failures require clusters centred on
the worst atoms, and the checks should keep running while the student is retrained. (3) Retraining converges the
student to the teacher, not to DFT; where the foundation model itself errs, as on under-coordinated Si at 3000~K, the
teacher must be validated and fine-tuned on DFT in the failing environments.

The approach has limits. Capacity and speed were measured on one GPU and on two GPUs of one node. The repair was
demonstrated with three seeds at $1.2\times10^{4}$ atoms and once at $1.03\times10^{6}$ atoms, and the two water
states with one seed each. Because the clusters carry no stress, retraining on them raised a small pressure bias in
ambient water (from 53 to 88--139~bar), so stress-carrying periodic cells should accompany the clusters where the
pressure matters. The students are short-ranged, which costs 1.2--2.1\% of the water density, and neither teacher nor
students contain explicit electrostatics. The throughput ratio to NEP89 refers to GPUMD~5.8 on the same GPU and will
change as descriptor-based implementations improve, whereas the ceiling of the exact equivariant model moves only if
its operations are reorganized.

%% ====================================================================================================================
%% Figures: r3c moved each figure* block into the text after the paragraph of its first mention

%% ====================================================================================================================
\begin{acknowledgement}
This research was supported under grant CFP04-CF-019. L.S. acknowledges financial support from the Singapore Ministry
of Education Tier 2 grant A-8001872-00-00. National University of Singapore Information Technology provided
computational resources.
\end{acknowledgement}

\section*{Data and Software Availability}
The training sets, DFT reference frames, records of all supervised runs and evaluation data are publicly available on
the Hugging Face Hub at \url{https://huggingface.co/datasets/shiqiao123/mace-whole-system-supervision-data}, and the
student models, including those retrained in the experiments, at
\url{https://huggingface.co/shiqiao123/mace-whole-system-supervision-models}. The code, including the three-pass
evaluation, the LAMMPS ML-IAP interface and patch, the distillation and supervision drivers, the benchmark scripts and
the input files, is publicly available at \url{https://github.com/ShiQiaoL/mace-million}; a source archive of the
release used here is included in the data repository.

\section*{Author Contributions}
Conception and design: X.L., Y.W., L.S.; data acquisition, computational investigation, formal analysis, visualization
and original draft: X.L.; methodology and mathematical analysis: X.L., Y.W.; supervision and critical revision: Y.W.,
L.S.; funding acquisition and project administration: L.S. All authors approved the final manuscript.

\section*{Notes}
The authors declare no competing financial interest.

\begin{suppinfo}\sloppy
Supplementary Notes 1--5 (exactness tests and two-GPU runs; operation count; changes to LAMMPS; cluster labels,
supervision experiments and hard cases; comparison with Symmetrix-XL), Supplementary Figures S1--S12 and
Supplementary Tables S1--S18 are appended to this preprint after the reference list. Supplementary Movies S1--S3
(MP4): the 1.03-million-atom SiO$_2$ check; the O--O contact and its repair; the supervised 1-ns run of
$9.83\times10^{5}$ water atoms.
\end{suppinfo}

\bibliography{refs}

% arXiv version: Supporting Information (compiled separately from arxiv/supplementary_arxiv.tex) appended as pages
\clearpage


\section*{Supplementary material}

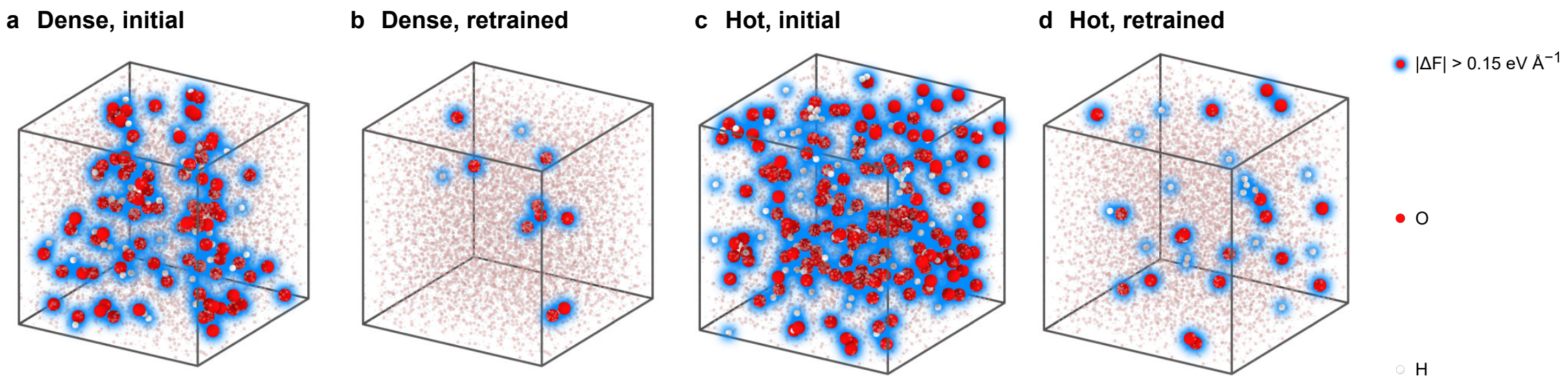


Figure S10: **Force deviation in the two water hard cases.** The teacher's own NVT frame at 30 ps of (a, b) compressed water (1.17 g cm$^{-3}$, 300 K, 47.2 Å box) and (c, d) water at 450 K (49.7 Å box), 12,288 atoms each, evaluated with (a, c) the initial student and (b, d) the final student of the worst-atom arm. No atom of these frames exceeds the 0.5 eV Å$^{-1}$ flag threshold except one in (c); atoms whose deviation from the teacher exceeds 0.15 eV Å$^{-1}$ are therefore drawn enlarged, in element colours, with the blue halo, and all other atoms small and pale. Their number falls from 149 to 11 (dense) and from 262 to 36 (hot) with retraining, and the component RMSE over all atoms from 36.8 to 25.5 and from 40.4 to 29.5 meV Å$^{-1}$: the hard cases produce a broad shoulder of moderate errors rather than a failure tail, and retraining lowers it throughout the box. Same orthographic view and scale in all panels. OVITO renders.

three repeats (spread below 1%). Device memory is the whole-process peak from `nvidia-smi` sampled every 0.5 s (a sampled peak, decimal GB). Symmetrix-XL was run in its `profile speed` and `profile capacity` modes with a single rank and without domain decomposition (identical throughput and memory within 1%), and in its tiled fixed-workspace mode (`mpi_message_passing`, `allow_fixed_workspace yes`). Force deviations are against the fp64 e3nn reference of the 192-atom water cell (ASE calculator).

*Node-to-node variation.* The throughput of our evaluation at 983,040 atoms was $2.3 \times 10^5$ atom-steps s$^{-1}$ on the node used for main-text Fig. 6a (24 September 2026; chunks of 32,768 and 65,536 receivers) and 1.8–$2.0 \times 10^5$ on the node of this comparison, with the same code, model file and settings; the per-call log attributes the difference to the model itself (the LAMMPS-side gap between calls is 5 ms), and the GPU ran at 1.9 GHz and 420 W (limit 600 W) in both cases. Within one node the measurements are reproducible to 1%. A timed breakdown of the force call on a third node (983,040 atoms, CUDA graphs, chunks of 8,192 to 65,536 receivers) gives 5.27, 4.97, 4.80 and 4.86 s per call, of which pass B takes 3.0–3.2 s and pass C 1.4–1.6 s; pass A takes 0.33–0.41 s, almost all of it copies into the static graph buffers, and the construction of the per-chunk local graphs only 6–25 ms. Building the local graphs of all chunks in a few batched passes (the default since this revision; results identical to $2 \times 10^{-6}$ eV Å$^{-1}$) changes the call time by less than 3% and halves the memory of the stored topologies. At $7.9 \times 10^6$ atoms with chunks of 8,192 receivers (960 chunks) one rank needs 55 s per call, eight times the 983,040-atom call with the same chunk size, of which 0.13 s is graph construction: the cost of small chunks is the lower efficiency of the kernels, not host work. On that node, running 983,040 atoms with chunks of 32,768 receivers without CUDA graphs was 8% faster than with them ($2.2 \times 10^5$ against $2.0 \times 10^5$ atom-steps s$^{-1}$) and used 64 instead of 77 GB; the graphs help at small chunks and on the slower nodes. Across nodes and these settings the throughput of the exact model at 983,040 atoms lies between 1.8 and $2.3 \times 10^5$ atom-steps s$^{-1}$.

# Supplementary Tables

Energy errors against DFT on the held-out frames are shown in Figure S12.

Table S12: Compressed water (12,288 atoms, 1.17 $\mathrm{g\,cm^{-3}}$, 300 K): pressure of frozen models in their own NVT dynamics, three independent runs of 20 ps per model. Run $k$ of every model starts from the same configuration of the teacher's reference trajectory (at $10k$ ps) with the same velocity seed. $\langle P\rangle$: mean pressure after 5 ps; SE: standard error from 1-ps block means; drift: slope of a linear fit over the same window; $\Delta P$: deviation from the teacher's run with the same start, with the error $\sqrt{\mathrm{SE}^2_{\mathrm{model}} + \mathrm{SE}^2_{\mathrm{teacher}}}$; mean: mean deviation over the three runs ± its standard error.

| model | run | $\langle P\rangle$ (kbar) | SE (bar) | drift (bar $\mathrm{ps^{-1}}$) | $\langle T\rangle$ (K) | $\Delta P$ (bar) |
|---|---|---|---|---|---|---|
| teacher | 1 | 20.47 | 34 | +9 | 300.1 | – |
| | 2 | 20.25 | 36 | −2 | 300.2 | – |
| | 3 | 20.48 | 35 | −20 | 300.4 | – |
| initial student (no retraining) | 1 | 18.98 | 33 | −3 | 300.2 | $-1498 \pm 47$ |
| | 2 | 19.09 | 24 | +2 | 299.7 | $-1165 \pm 44$ |
| | 3 | 18.91 | 26 | +9 | 299.9 | $-1573 \pm 44$ |
| | mean | | | | | $-1412 \pm 125$ |
| retrained, worst-atom clusters | 1 | 20.36 | 45 | +23 | 300.2 | $-112 \pm 56$ |
| | 2 | 20.66 | 29 | +1 | 299.9 | $+406 \pm 47$ |
| | 3 | 20.25 | 28 | −1 | 299.9 | $-233 \pm 45$ |
| | mean | | | | | $+20 \pm 196$ |
| retrained, random clusters | 1 | 20.59 | 21 | +11 | 300.2 | $+114 \pm 40$ |
| | 2 | 20.36 | 37 | −24 | 299.7 | $+106 \pm 52$ |
| | 3 | 20.38 | 18 | −10 | 300.0 | $-107 \pm 39$ |
| | mean | | | | | $+38 \pm 72$ |

Table S13: Supervised segments of $1.92 \times 10^5$ water atoms (2,000 steps each, NPT 300 K): force RMSE of the student against the teacher on the worst checked frame (meV $\mathrm{Å^{-1}}$) and virial-pressure deviation (bar).

| | force RMSE | | | pressure deviation | | |
|---|---|---|---|---|---|---|
| segment | 1 | 2 | 3 | 1 | 2 | 3 |
| retrained, 7-Å clusters | 23.8 | 21.0 | 19.8 | 117 | 112 | 5 |
| retrained, 13-Å clusters | 23.8 | 20.9 | 19.9 | 117 | 62 | 59 |
| no retraining (control) | 23.8 | 21.8 | 21.4 | 117 | 25 | 32 |
| density ($\mathrm{g\,cm^{-3}}$, all runs) | 0.931 | 0.906 | 0.903–0.904 | | | |

# Supplementary Movies

**Movie S1. Whole-system checks during continuous MD of the $1.03 \times 10^6$-atom $SiO_2$ runs at 3000 K.** Both runs continue for 1 ps (2 to 3 ps) from their restart files at 2 ps: left with the initial student (fixed run), right with the student retrained once after the first segment. Shown is an 80 × 80 × 20 Å window at the box centre (about 7,300 atoms that stay in it for the whole picosecond), dumped every 10 fs and drawn every 5 fs (linear interpolation). Every 50 fs the teacher evaluates the whole box; atoms whose force deviates from the teacher by more than 0.5 eV $\mathrm{Å^{-1}}$ at the latest check are drawn enlarged in their element colour (Si grey, O red) with a blue halo, all other atoms pale. Whole box, 21 checks: 987–1,212 flagged atoms with the fixed student and 359–485 with the retrained one (1–20 and 0–8 in the window). The check at 2 ps uses the model that drives the continuation, so the retrained run starts at 359 rather than the 1,121 of the production check with the initial student; at 3 ps both runs reproduce the production checks exactly (1,094 and 451). The time stamp is the simulated time; the curve below gives the flagged atoms of the whole box at each check (grey: fixed, green: retrained; checks already passed drawn solid); the spheres at the lower right are the flagged Si and O as drawn. Viewed nearly along $z$, window edges parallel to the frame. OVITO (Tachyon, ambient occlusion).

Table S14: Record of the supervised $SiO_2$ segments (NPT, 3000 K, 2 ps each, two checked frames at 1 and 2 ps into the segment), reconstructed from the run directories. Student: model that drove the segment (c0: initial; c$k$: after $k$ retrainings). Worst checked frame: force RMSE (meV Å$^{-1}$), fraction of atoms above 0.5 eV Å$^{-1}$ ($10^{-4}$) and $|\Delta P|$ (bar) against the teacher. Thresholds: 90 meV Å$^{-1}$, $10^{-5}$ ($0.1 \times 10^{-4}$) and 3,000 bar; the fixed runs check without retraining. Clusters: 13-Å clusters labelled for the retraining after the segment. Stage durations (s, wall time): MD including LAMMPS start-up, teacher and student checks, cluster labels, fine-tuning (40 epochs), and the whole segment including the model conversion (9–13 s). Every segment was a first attempt and started from the end of the previous one; no safety stop fired.

| | | | worst checked frame | | | | | wall time (s) | | | | |
|---|---|---|---|---|---|---|---|---|---|---|---|---|
| atoms, run | segment (ps) | student | RMSE | fraction | $\|\Delta P\|$ | criteria fired | clusters | MD | checks | labels | fine-tuning | total |
| $1.03 \times 10^6$, retrained | 0–2 | c0 | 74.0 | 10.9 | 879 | fraction | 8 | 552 | 146 | 20 | 217 | 946 |
| | 2–4 | c1 | 67.6 | 4.4 | 47 | fraction | 8 | 472 | 147 | 18 | 226 | 874 |
| | 4–6 | c2 | 218.1 | 3.5 | 224 | RMSE, fraction | 8 | 454 | 153 | 19 | 231 | 868 |
| | 6–8 | c3 | 97.8 | 3.1 | 249 | RMSE, fraction | 8 | 455 | 153 | 19 | 238 | 876 |
| | 8–10 | c4 | 65.9 | 2.9 | 319 | fraction | 8 | 454 | 156 | 18 | 242 | 881 |
| $1.03 \times 10^6$, fixed | 0–2 | c0 | 74.0 | 11.0 | 879 | fraction | – | 539 | 142 | – | – | 691 |
| | 2–4 | c0 | 74.5 | 10.6 | 815 | fraction | – | 464 | 137 | – | – | 613 |
| | 4–6 | c0 | 74.3 | 10.9 | 825 | fraction | – | 446 | 136 | – | – | 593 |
| | 6–8 | c0 | 74.3 | 11.0 | 827 | fraction | – | 449 | 150 | – | – | 610 |
| | 8–10 | c0 | 74.2 | 11.4 | 830 | fraction | – | 446 | 149 | – | – | 609 |
| 12,288, retrained (worst atoms, seed 1) | 0–2 | c0 | 77.4 | 14.6 | 908 | fraction | 8 | 69 | 84 | 5 | 224 | 392 |
| | 2–4 | c1 | 67.3 | 7.3 | 32 | fraction | 8 | 69 | 83 | 3 | 228 | 395 |
| | 4–6 | c2 | 64.2 | 6.5 | 246 | fraction | 8 | 68 | 83 | 3 | 233 | 399 |
| | 6–8 | c3 | 61.7 | 1.6 | 142 | fraction | 8 | 69 | 84 | 3 | 241 | 409 |
| | 8–10 | c4 | 63.6 | 1.6 | 853 | fraction | 8 | 69 | 85 | 3 | 245 | 413 |
| 12,288, fixed (seed 1) | 0–2 | c0 | 77.4 | 13.8 | 908 | fraction | – | 70 | 84 | – | – | 163 |
| | 2–4 | c0 | 73.7 | 16.3 | 880 | fraction | – | 69 | 82 | – | – | 160 |
| | 4–6 | c0 | 73.3 | 8.1 | 793 | fraction | – | 69 | 87 | – | – | 165 |
| | 6–8 | c0 | 73.3 | 13.8 | 963 | fraction | – | 68 | 84 | – | – | 164 |
| | 8–10 | c0 | 75.3 | 11.4 | 863 | fraction | – | 68 | 83 | – | – | 161 |

Table S15: Symmetrix-XL 0.1.1 and this work on the same H200, MACE-MPA-0, liquid water, fp32. Throughput in atom-steps s$^{-1}$; memory is the whole-process device peak.

| | Symmetrix-XL | this work |
|---|---|---|
| force deviation, 192 atoms, fp32 (max / RMS, eV Å$^{-1}$) | $1.6 \times 10^{-4}$ / $4.7 \times 10^{-5}$ | $2.1 \times 10^{-5}$ / $5.7 \times 10^{-6}$ |
| force deviation, 192 atoms, fp64 | $6.1 \times 10^{-5}$ / $2.9 \times 10^{-5}$ | $1 \times 10^{-14}$ / $3 \times 10^{-15}$ |
| 12,288 atoms | $2.17 \times 10^5$, 2.0 GB | $1.99 \times 10^5$, 14 GB |
| 98,304 atoms | $2.12 \times 10^5$, 8.7 GB | $2.09 \times 10^5$, 64 GB |
| 983,040 atoms, chunks of 65,536 | $2.10 \times 10^5$, 77 GB | $2.02 \times 10^5$, 136 GB |
| 983,040 atoms, chunks of 32,768 | | $1.76 \times 10^5$, 81 GB |
| $7.86 \times 10^6$ atoms, one rank | tiled: $2.06 \times 10^5$, 95 GB; other modes: illegal address | $1.85 \times 10^5$, 128 GB[a] |
| $9.83 \times 10^6$ atoms, two ranks per GPU | tiled: $1.79 \times 10^5$, 122 GB | out of memory |
| $1.18 \times 10^7$ atoms, two ranks per GPU | out of memory | – |
| largest system in LAMMPS, one H200 | $9.8 \times 10^6$ | $9.0 \times 10^6$ (32-bit neighbour list) |
| largest system standalone, one H200 | not measured | $2.86 \times 10^7$ |
| stock ML-IAP + cuEquivariance, for scale | $6.0 \times 10^4$ at 12,288 atoms; $6.5 \times 10^4$ at 41,472; out of memory at 86,016 (300-step runs; a 3-step evaluation fits, main-text Fig. 2c) | |

[a] Chunks of 16,384 receivers without CUDA graphs; with graphs the static buffers do not fit at this size.

**Movie S2. The O–O contact forming in the supervised $1.03 \times 10^6$-atom $SiO_2$ run.** Continuous replay of the segment from 4 to 6 ps (restart at 4 ps, the student after two retraining rounds that drove this segment, unchanged settings), 5 fs per movie frame. Shown: the atoms within 8 Å of the two contact pairs and within 3 Å of the plane through pair 1 and the line joining the pairs (67 atoms), viewed along the normal of that plane; the four contact O atoms carry a blue halo. Left: the forces of the student that drives the dynamics on the four contact atoms; right: the same configurations with the teacher forces, evaluated every 100 fs (the movie pauses at each evaluation). Arrows: 1 Å per 40 eV Å$^{-1}$, at most 5 Å long. Curves: the two O–O distances and the repulsion on each pair (force along the O–O axis, mean of the two atoms, positive when it pushes them apart). In the replay the pairs close to 0.68 Å (5.66 ps) and 0.70 Å (5.93 ps), against 0.815 and 0.822 Å in the production frame at 6 ps; the student's repulsion stays between $-27$ and $+104$ eV Å$^{-1}$, while the teacher repels with up to 155 (pair 1) and 364 eV Å$^{-1}$ (pair 2). The time stamp is the simulated time; the teacher arrows appear only on the evaluated configurations. The distance scan of the contact frame with all four models is Fig. S9.

**Movie S3. Water with $9.83 \times 10^5$ atoms after 1 ns of supervised production MD.** Left: 2 ps of continuous dynamics (NPT, 300 K, 1 atm, 1 fs steps) with the distilled student that

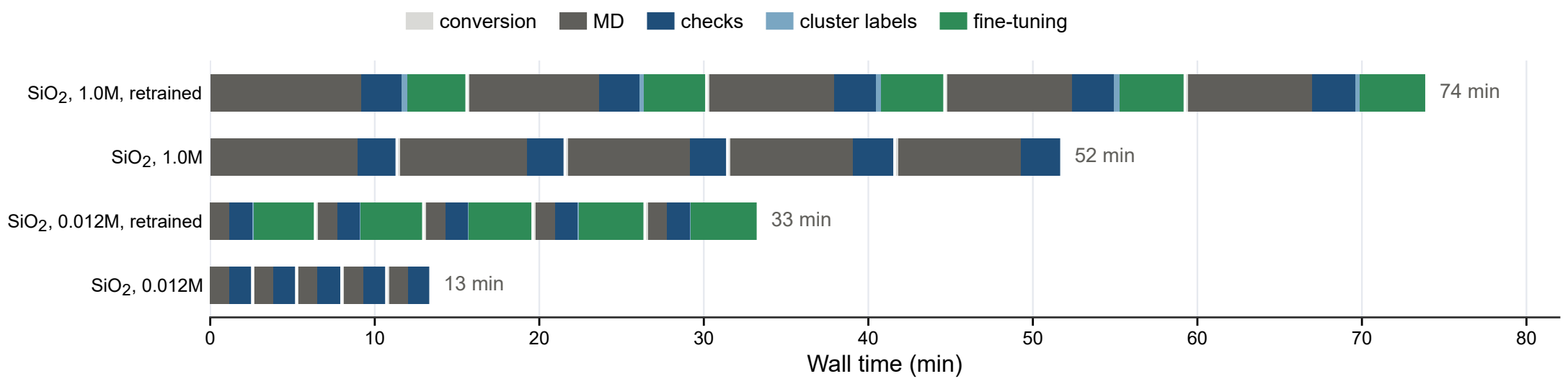


Figure S11: Wall-clock timeline of the supervised $SiO_2$ runs of Table S14, by stage, reconstructed from the modification times of the files that each stage writes; 1.0M: $1.03 \times 10^6$ atoms, 0.012M: 12,288 atoms. Each run is a sequence of segments, separated by white gaps; every segment starts from the end of the previous one, and no segment was repeated; the runs without 'retrained' check without retraining. The total wall time is given at the end of each bar. The share of each stage is shown in main-text Fig. 5f; relative to the MD time, the checks of the $1.03 \times 10^6$-atom runs add about 30%.

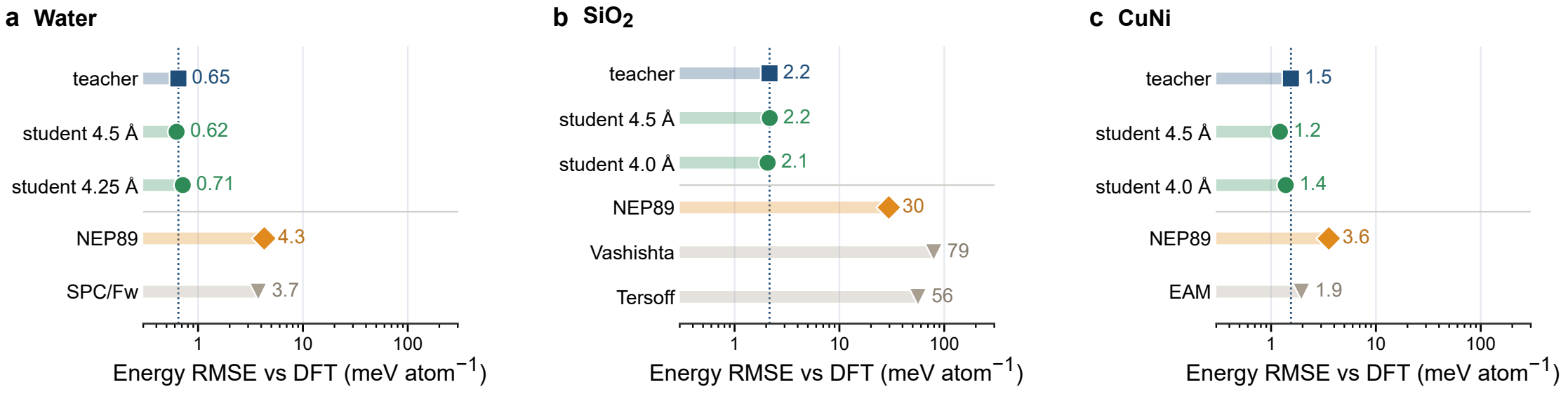


Figure S12: **Energy errors against DFT on held-out frames.** Energy RMSE against DFT, with the mean offset of each model removed, for (a) water and (b) $SiO_2$ (20 frames each) and (c) CuNi (4 frames); same models, frames and order as the force errors of main-text Fig. 6d–f. Rows: the MACE-MPA-0 teacher, distilled students with the cutoffs indicated, NEP89, and classical force fields (SPC/Fw; Vashishta and Tersoff; EAM). Dotted lines: the teacher's error. Logarithmic axes. The students are within 0.4 meV $atom^{-1}$ of the teacher in all three systems (water 0.62 and 0.71 against 0.65; $SiO_2$ 2.16 and 2.07 against 2.16; CuNi 1.21 and 1.38 against 1.55 meV $atom^{-1}$); NEP89 gives 4.3, 30 and 3.6 meV $atom^{-1}$.

drove the whole 1-ns run, started from the restart file at 1 ns; the 382 molecules that start in a $36 \times 36 \times 12$ Å region at the box centre (O red, H white), dumped every 20 fs and shown at 6.7 fs per movie frame (linear interpolation between dumps). Right: the record of the 1-ns run: number of atoms deviating from the teacher by more than 0.5 eV $Å^{-1}$ in each of the 100 whole-system checks (every 10 ps; at most three of $9.83 \times 10^5$), and the density at each check (0.905 to 0.912 $g\,cm^{-3}$); the student was never retrained. The time stamp counts from the end of the 1-ns run; the shaded band in both plots marks the 2 ps of the movie; scale bar 10 Å. Viewed nearly along $z$, with the region edges parallel to the frame. OVITO (Tachyon, ambient occlusion).

Table S16: Software and hardware.

| | |
|---|---|
| GPU | NVIDIA H200, 141 GB (143,771 MiB), driver 575.57.08, one GPU per run |
| CUDA / PyTorch | CUDA 12.4, PyTorch 2.5.1, Triton 3.1.0 |
| MACE / e3nn / cuEquivariance | MACE 0.3.15, e3nn 0.4.4, cuEquivariance 0.9.1 |
| LAMMPS (MLIP runs) | 11 Feb 2026 (commit 5c6749b), Kokkos 5.0.2, ML-IAP unified, Python 3.11.14 |
| LAMMPS (classical runs) | 22 Jul 2025 (official PyPI wheel, CPU) |
| NEP89 | `nep89_20250409`, GPUMD 5.8 |
| DFT | VASP 6.5.0 (GPU build, nvhpc 24.5) |
| ASE / NumPy | ASE 3.28.0, NumPy 2.0.1 |

Table S17: DFT settings (compatible with the Materials Project / MPtrj data on which the teacher was trained).

| | |
|---|---|
| functional | PBE |
| pseudopotentials | PAW PBE_54 with Materials Project choices (H, O, Si, Cu_pv, Ni_pv, ...) |
| plane-wave cutoff | 520 eV (`PREC = Accurate`, `LASPH = .TRUE.`, `LREAL = .FALSE.`) |
| smearing | Gaussian, 0.05 eV |
| electronic convergence | $10^{-5}$ eV (`ALGO = Fast`) |
| spin | polarised for systems containing magnetic elements (CuNi; initial moments 0.6 $\mu_B$) |
| *k*-points | $\Gamma$ (water, $SiO_2$, 192 atoms); $2 \times 2 \times 2$ $\Gamma$-centred (CuNi, 108 atoms) |

Table S18: Classical force fields.

| | |
|---|---|
| water | SPC/Fw (Wu, Tepper, Voth 2006): $q_{\mathrm{O}} = -0.82e$, $q_{\mathrm{H}} = 0.41e$, O–O LJ $\varepsilon = 0.1554$ kcal mol$^{-1}$, $\sigma = 3.1655$ Å;<br>harmonic bond $k = 1059.162$ kcal mol$^{-1}$ Å$^{-2}$, $r_0 = 1.012$ Å; angle $k = 75.90$ kcal mol$^{-1}$ rad$^{-2}$, $\theta_0 = 113.24°$; Ewa |
| $SiO_2$ | Vashishta 1990 (`SiO.1990.vashishta`); Tersoff/Munetoh (`SiO.tersoff`), LAMMPS distribution |
| CuNi | EAM, Onat and Durukanoglu 2014 (`CuNi.eam.alloy`), LAMMPS distribution |



\end{document}